\documentclass{emulateapj}
\usepackage{natbib}
\usepackage{graphicx}
\usepackage{times}
\usepackage{xspace} % to sort out spaces after \newcommand symbols/expressions
\usepackage{booktabs} % for space between horizontal lines in table 1

\newcommand {\apgt} {\ {\raise-.5ex\hbox{$\buildrel>\over\sim$}}\ }
\newcommand {\aple} {\ {\raise-.5ex\hbox{$\buildrel<\over\sim$}}\ }

\newcommand {\chandra} {{\it Chandra}\xspace}
\newcommand {\xmm} {{\it XMM-Newton}\xspace}
\newcommand {\nustar} {{\it NuSTAR}\xspace}
\newcommand {\spitzer} {{\it Spitzer}\xspace}

\newcommand {\sixum} {$6$~$\mu$m\xspace}
\newcommand {\feka} {Fe~K$\alpha$\xspace}

\newcommand{\oiii}{\mbox{[\ion{O}{3}]}\xspace}

\newcommand{\typei}{Type 1\xspace}
\newcommand{\typeii}{Type 2\xspace}

\newcommand {\ergpersec} {erg~s$^{-1}$\xspace}
\newcommand {\fluxunit} {erg~s$^{-1}$~cm$^{-2}$\xspace}
\newcommand {\nhunit} {cm~$^{-2}$\xspace}

\newcommand {\sdssa} {SDSS~J0011+0056\xspace}
\newcommand {\sdssb} {SDSS~J0056+0032\xspace}
\newcommand {\sdssc} {SDSS~J1157+6003\xspace}

\begin{document}
%%%%%%%%%%%%%%%%%%%%%%%%%%%%%%%%%%%%%%%%%%%%%%%%%%%%%%%%%%%%%%%%%%%%%%

%%%%%%%%%%%%%%%%%%%%%%%%%%%%%%%%%%%%%%%%%%%%%%%%%%%%%%%%%%%%%%%%%%%%%%
\title{\nustar Observations of Heavily Obscured Quasars at $z\sim0.5$}
%%%%%%%%%%%%%%%%%%%%%%%%%%%%%%%%%%%%%%%%%%%%%%%%%%%%%%%%%%%%%%%%%%%%%%

%%%%%%%%%%%%%%%%%%%%%%%%%%%%%%%%%%%%%%%%%%%%%%%%%%%%%%%%%%%%%%%%%%%%%%
\author{G.~B.~Lansbury\altaffilmark{1},
  D.~M.~Alexander\altaffilmark{1}, 
% joint third author tier:
A.~Del~Moro\altaffilmark{1}, P.~Gandhi\altaffilmark{1},
% next tier:
R.~J.~Assef\altaffilmark{2}, D.~Stern\altaffilmark{3},
% last tier:
J.~Aird\altaffilmark{1}, D.~R.~Ballantyne\altaffilmark{4}, M.~Balokovi\'c\altaffilmark{5},
F.~E.~Bauer\altaffilmark{6,7}, S.~E.~Boggs\altaffilmark{8},
W.~N.~Brandt\altaffilmark{9,10}, F.~E.~Christensen\altaffilmark{11}, W.~W.~Craig\altaffilmark{11,12}, M.~Elvis\altaffilmark{13}, B.~W.~Grefenstette\altaffilmark{5},
C.~J.~Hailey\altaffilmark{14}, F.~A.~Harrison\altaffilmark{5},
R.~C.~Hickox\altaffilmark{15}, M.~Koss\altaffilmark{16},
S.~M.~LaMassa\altaffilmark{17}, B.~Luo\altaffilmark{9,10}, J.~R.~Mullaney\altaffilmark{1}, S.~H.~Teng\altaffilmark{18}, C.~M.~Urry\altaffilmark{17}, W.~W.~Zhang\altaffilmark{19}}
%%%%%%%%%%%%%%%%%%%%%%%%%%%%%%%%%%%%%%%%%%%%%%%%%%%%%%%%%%%%%%%%%%%%%%

%%%%%%%%%%%%%%%%%%%%%%%%%%%%%%%%%%%%%%%%%%%%%%%%%%%%%%%%%%%%%%%%%%%%%%
\affil{$^1$Department of Physics, University of Durham, South Road,
  Durham DH1 3LE, UK; g.b.lansbury@durham.ac.uk}
% Assef:
\affil{$^2$N\'ucleo de Astronom\'ia de la Facultad de Ingenier\'ia,
  Universidad Diego Portales, Av. Ej\'ercito Libertador 441, Santiago,
  Chile}
% Stern:
\affil{$^3$Jet Propulsion Laboratory, California Institute of Technology, 4800 Oak Grove Drive, Mail Stop 169-221, Pasadena, CA 91109, USA}
% Ballantyne:
\affil{$^4$Center for Relativistic Astrophysics, School of Physics, Georgia Institute of Technology, Atlanta, GA 30332, USA}
% Balokovic:
\affil{$^5$Cahill Center for Astrophysics, 1216 East California
  Boulevard, California Institute of Technology, Pasadena, CA 91125,
  USA}
% Bauer:
\affil{$^6$Instituto de Astrof\'{\i}sica, Facultad de F\'{i}sica, Pontificia Universidad Catlica de Chile, 306, Santiago 22, Chile}
\affil{$^7$Space Science Institute, 4750 Walnut Street, Suite 205,
  Boulder, Colorado 80301, USA}
% Boggs:
\affil{$^8$Space Sciences Laboratory, University of California,
  Berkeley, CA 94720, USA}
% Brandt & Luo:
\affil{$^9$Department of Astronomy and Astrophysics, 525 Davey Lab,
  The Pennsylvania State University, University Park, PA 16802, USA}
\affil{$^{10}$Institute for Gravitation and the Cosmos, The Pennsylvania
  State University, University Park, PA 16802, USA}
% Christensen & Craig_1:
\affil{$^{11}$DTU Space-National Space Institute, Technical University of
  Denmark, Elektrovej 327, DK-2800 Lyngby, Denmark}
% Craig_2:
\affil{$^{12}$Lawrence Livermore National Laboratory, Livermore, CA
  94550, USA}
% Elvis:
\affil{$^{13}$Harvard-Smithsonian Center for Astrophysics, 60 Garden
  Street, Cambridge, MA 02138, USA}
% Hailey:
\affil{$^{14}$Columbia Astrophysics Laboratory, 550 W 120th Street,
  Columbia University, NY 10027, USA}
% Hickox:
\affil{$^{15}$Department of Physics and Astronomy, Dartmouth College, 6127 Wilder Laboratory, Hanover, NH 03755, USA}
% Koss:
\affil{$^{16}$Institute for Astronomy, Department of Physics, ETH Zurich, Wolfgang-Pauli-Strasse 27, CH-8093 Zurich, Switzerland}
%LaMassa:
\affil{$^{17}$Yale Center for Astronomy and Astrophysics, Physics Department, Yale University, PO Box 208120, New Haven, CT 06520-8120, USA}
%Teng:
\affil{$^{18}$Observational Cosmology Laboratory, NASA Goddard Space
  Flight Center, Greenbelt, MD 20771, USA}
%Zhang:
\affil{$^{19}$NASA Goddard Space Flight Center, Greenbelt, MD 20771, USA}
%%%%%%%%%%%%%%%%%%%%%%%%%%%%%%%%%%%%%%%%%%%%%%%%%%%%%%%%%%%%%%%%%%%%%%
\shorttitle{\nustar Observations of Heavily Obscured Quasars at $z\sim0.5$}

\shortauthors{Lansbury et al.}
%%%%%%%%%%%%%%%%%%%%%%%%%%%%%%%%%%%%%%%%%%%%%%%%%%%%%%%%%%%%%%%%%%%%%%

%%%%%%%%%%%%%%%%%%%%%%%%%%%%%%%%%%%%%%%%%%%%%%%%%%%%%%%%%%%%%%%%%%%%%%
%%%%%%%%%%%%%%%%%%%%%%%%%%%%%%%%%%%%%%%%%%%%%%%%%%%%%%%%%%%%%%%%%%%%%%
\begin{abstract}
We present \nustar hard X-ray %($3$--$79$~keV) 
observations of three \typeii quasars at
$z\approx0.4$--$0.5$, optically selected from the Sloan Digital Sky
Survey (SDSS). Although the quasars show evidence for being heavily
obscured Compton-thick systems on the basis of the $2$--$10$~keV to \oiii
luminosity ratio and multiwavelength diagnostics, their X-ray absorbing column densities
($N_{\rm H}$) are poorly known.
In this analysis: (1) we study X-ray emission at $>10$~keV, where X-rays from the
central black hole are relatively unabsorbed, in order to better constrain
$N_{\rm H}$; (2) we further characterize the physical properties of the sources through broad-band
near-UV to mid-IR spectral energy distribution (SED) analyses.
One of the quasars is detected with \nustar at $>8$~keV with a
no-source probability of $<0.1\%$, and its X-ray band ratio suggests 
near Compton-thick absorption with $N_{\rm H} \gtrsim 5 \times 
10^{23}$~\nhunit.
The other two quasars are undetected, and have low X-ray to mid-IR luminosity
ratios in both the low energy ($2$--$10$~keV) and high energy ($10$--$40$~keV) X-ray
regimes that are consistent with extreme,
Compton-thick absorption ($N_{\rm H}\gtrsim 10^{24}$~\nhunit).
We find that for quasars at $z\sim 0.5$, \nustar provides a significant improvement compared to
lower energy ($<10$~keV) \chandra and \xmm observations alone, as higher column
densities can now be directly constrained.
\end{abstract}
%%%%%%%%%%%%%%%%%%%%%%%%%%%%%%%%%%%%%%%%%%%%%%%%%%%%%%%%%%%%%%%%%%%%%%

%%%%%%%%%%%%%%%%%%%%%%%%%%%%%%%%%%%%%%%%%%%%%%%%%%%%%%%%%%%%%%%%%%%%%%
%%%%%%%%%%%%%%%%%%%%%%%%%%%%%%%%%%%%%%%%%%%%%%%%%%%%%%%%%%%%%%%%%%%%%%

%%%%%%%%%%%%%%%%%%%%%%%%%%%%%%%%%%%%%%%%%%%%%%%%%%%%%%%%%%%%%%%%%%%%%%
\section{Introduction}
\label{Introduction}
%%%%%%%%%%%%%%%%%%%%%%%%%%%%%%%%%%%%%%%%%%%%%%%%%%%%%%%%%%%%%%%%%%%%%%

Quasars are the sites of the most rapid black hole growth in the
universe \citep{Salpeter64,Soltan82}. They represent the luminous
end of the active galactic nucleus (AGN) population, often outshining
their host galaxies. The first unobscured (`\typei') quasars were
discovered over 50 years ago \citep{Schmidt63, Hazard63}, and
more than one hundred thousand have now been spectroscopically
identified (e.g., \citealt{VeronCetty10, Paris12}).  For obscured
(`\typeii') quasars\footnote{We define \typeii quasars as AGN with
$L_{\rm 2-10keV} \geq 10^{44}$~\ergpersec, X-ray absorbing column
densities $N_{\rm H}>10^{22}$~\nhunit, and optical spectra that
show narrow line emission without broad (H$\rm \alpha$ or H$\rm
\beta$) components.  This $L_{\rm 2-10keV}$ threshold is consistent
with: (1) the classical optical quasar definition, $M_{B}\leq -23$,
when the $\alpha _{\rm OX}$ relation of \citet{Steffen06} and the
composite quasar spectrum of \citet{VandenBerk01} are assumed; (2)
the $L_{\rm X,*}$ value derived by \citet{Hasinger05} for unobscured
AGN.} the situation is not as advanced.  Similar to the early \typei
quasars, \typeii quasars were initially identified from radio
selection (e.g., \citealt{Minkowski60}), and over the following
decades several hundred powerful `radio galaxies' (as such
radio-selected \typeii\ quasars are typically called) were identified
(for reviews, see \citealt{McCarthy93, Miley08}).  However, it is
only in the past decade that radio-quiet \typeii quasars have been
found in large numbers.  Such sources are generally identified on
the basis of either their relatively hard X-ray spectral slopes (e.g., \citealt{Norman02,
Stern02}), optical spectral features (e.g., \citealt{Steidel02, Zakamska03}),
or mid-infrared (mid-IR) colors (e.g., \citealt{Lacy04, Stern05}).  Importantly,
mid-IR color selection of \typeii quasars using the all-sky
{\it Wide-Field Infrared Survey Explorer} ({\it WISE}; \citealt{Wright10})
survey identifies several million \typeii quasars, roughly down to
the bolometric luminosity of the primary Sloan Digital Sky Survey \citep[SDSS;][]{York00} \typei quasar
spectroscopic survey \citep{Stern12, Assef13, Donoso13}.

The exact nature of \typeii quasars is still under debate. A simple
extension of the orientation-driven unified model of AGN
\citep{Antonucci93, Urry95} to high luminosities can account for
their existence. However, there is also observational evidence for
an evolutionary link to \typei quasars \citep[e.g.,][]{Sanders88,
Hopkins08}.  The importance of \typeii quasars to the cosmic evolution
of AGN is further demonstrated by their requirement in models of
the cosmic X-ray background (CXB) \citep[e.g.,][]{Treister05,
Gilli07, Treister09}.  However, the observed X-ray properties of
\typeii quasars are poorly constrained at present. Consequently,
the column density ($N_{\rm H}$) distribution\footnote{X-rays emitted
from the immediate black hole environment are absorbed by circumnuclear
gas, and thus provide constraints on $N_{\rm H}$.} and
Compton-thick\footnote{Compton-thick absorption is that with $N_{\rm
H} \ge {\rm \sigma}_{\rm T}^{-1}\approx1.5\times 10^{24}$~\nhunit.}
fraction of quasars are poorly known, which has implications for
both AGN and CXB models \citep[e.g.,][]{Fabian08,Draper10}.

%(2) IDENTIFICATION OF TYPE 2 QUASARS:

To date, the largest sample of spectroscopically confirmed (radio-quiet)
\typeii quasars at $z\lesssim1$ is that of
\citet{Zakamska03} and \citet{Reyes08}.
\citet{Zakamska03} selected 291 \typeii quasars at redshift $0.2 \lesssim z
\lesssim 0.8$ from the SDSS based on
their optical properties: high \oiii $\lambda 5007$
line power and narrow emission lines.
\citet{Reyes08} used the same approach and more recent SDSS data to extend the sample to 887 objects.
%\typeii quasars have also been selected at radio, mid-infrared and X-ray wavelengths 
%\citep[e.g.,][]{
%% Radio:
%% X-ray:
%Norman02,Stern02,Barger03,Fiore03,Gandhi04a,
%% mid-IR:
%Lacy07,Eisenhardt12,Stern12,Assef13}. 
While X-ray selections of \typeii quasars at $\lesssim 10$~keV are biased against the most heavily obscured sources \citep[e.g.,][]{Maiolino98a},
\oiii emission is mostly produced on $\sim100$~pc scales and is thus
relatively unaffected by nuclear obscuration, allowing larger numbers
of the heavily obscured, X-ray faint objects to be found.
Following up \oiii selected, rather than X-ray selected, objects with X-ray observations thus gives a
less biased estimate of the $N_{\rm H}$ distribution
of AGN \citep[e.g.,][]{Risaliti99}.

%(3) MEASURING/IDENTIFYING HEAVY OBSCURATION:

The X-ray properties of the \citet{Zakamska03} and \citet{Reyes08}
\typeii quasar sample have been studied using \chandra and \xmm
observations \citep{Ptak06,Vignali06,Vignali10,Jia13}.
\citet{Vignali06,Vignali10} measured column
densities for a handful of sources through `direct' means (i.e., using
X-ray spectroscopic analysis). The highest column densities measured in
this manner were $N_{\rm H}\approx 3
\times 10^{23}$~\nhunit.
However, distant obscured quasars are X-ray weak and in most cases
direct constraints are not feasible. Instead, an `indirect' approach
to estimating column densities
can be used where the observed X-ray emission is compared with a proxy for intrinsic
AGN power \citep[e.g., the mid-IR continuum emission from hot dust or high-excitation
emission lines;][]{Bassani99,Lutz04,Heckman05,Alexander05b,Alexander08,Cappi06,Panessa06,Melendez08,Gandhi09,LaMassa09,LaMassa11,Gilli10,Goulding11}.
\citet{Vignali06,Vignali10} were limited to indirect absorption
constraints for the majority of their \typeii quasar sample, and found 
in every case that Compton-thick absorption (i.e., $N_{\rm H} > 1.5
\times 10^{24}$~\nhunit) is required to explain the X-ray
suppression in these sources.
To first order, there appears to be a bimodal $N_{\rm H}$
distribution for optically selected \typeii quasars, with $\sim 40\%$ having $N_{\rm
  H} = 10^{22}$--$3\times10^{23}$~\nhunit and $\sim 60\%$ being
Compton-thick.
This is interesting given that a 
continuous $N_{\rm H}$ distribution is measured for %optically and mid-IR selected
\typeii Seyferts \citep[e.g.,][]{Bassani99,Risaliti99,LaMassa09,LaMassa11}, although the differences may be reconciled by
considering the different methods used to estimate $N_{\rm H}$ \citep{LaMassa11}.
To better constrain the $N_{\rm H}$
distribution of \typeii quasars, more robust identifications of Compton-thick
absorption must be obtained through either:
(i) measurement of strong \feka emission,
with ${\rm EW}\geq1$~keV, which results from the \feka line being viewed in
reflection against a suppressed continuum \citep[e.g.,][]{Ghisellini94,Levenson02}; % Jia+13 have done this, but in an average sense
or (ii) measurement of high column densities through spectroscopic
analysis at high energies above the photoelectric
absorption cutoff (i.e., above observed-frame $8$~keV for $z\sim
0.5$ and $N_{\rm H}\sim10^{24}$~cm$^{-2}$), where X-ray emission is
relatively unabsorbed.

%(5) NUSTAR:

The recent launch of the {\it Nuclear Spectroscopic Telescope Array}
\citep[\nustar,][]{Harrison13} will see a breakthrough in our
understanding of heavily obscured AGN and the CXB population in general. 
\nustar is the first orbiting observatory with the ability to focus high
energy ($\gtrsim 10$~keV) X-rays using grazing incidence optics.
It provides a two orders of magnitude
improvement in sensitivity and over an order of magnitude improvement in
angular resolution over previous hard X-ray
observatories. 
The high energy range at which \nustar operates ($3$--$79$~keV) means that
the intrinsic, unabsorbed emission of AGN is observed for all but the
most heavily obscured, Compton-thick objects.
At $z \lesssim 1$, it is now possible to directly constrain column
densities an order of magnitude higher than those achievable with \chandra and
\xmm alone \citep[e.g.,][]{Luo13}. 

In this paper, we present exploratory \nustar observations of three
optically selected \typeii quasars at $z \approx 0.4$--$0.5$. All three have
been identified as Compton-thick candidates in previous studies
\citep{Vignali06,Vignali10,Jia13}. We use X-ray data from \nustar,
\chandra and \xmm, and near-UV to mid-IR data from other observatories to
determine the physical properties of the quasars. In particular, we use a combination of
direct and indirect methods to constrain the absorbing column densities.
The paper is organized as follows: our sample selection is
detailed in Section \ref{sample}; we describe the observations, data reduction and data analysis in Section
\ref{data}; %, including: \nustar data, photometry and source
%detection (Section \ref{nustar_data}); lower energy X-ray data
%(Section \ref{low_data}); broadband UV--mid-IR data and spectral energy
%distribution (SED) analysis (Section \ref{UVMIR_data}). 
our main results
regarding X-ray absorption constraints are presented in Section \ref{Results};
we summarize our main conclusions in Section \ref{summary}.
The cosmology adopted throughout this work is
$(\Omega_{M},\Omega_{\Lambda},h)=(0.27,0.73,0.71)$.
%%%%%%%%%%%%%%%%%%%%%%%%%%%%%%%%%%%%%%%%%%%%%%%%%%%%%%%%%%%%%%%%%%%%%%
\section{Sample Selection}
\label{sample}
%%%%%%%%%%%%%%%%%%%%%%%%%%%%%%%%%%%%%%%%%%%%%%%%%%%%%%%%%%%%%%%%%%%%%%

First, we selected objects at $z \approx 0.4$--$0.5$ from the \chandra and
\xmm studies of SDSS selected \typeii quasars by \citet{Vignali06,Vignali10} and
\citet{Jia13}. 
Although the objects have narrow H$\rm \beta$ line emission, the H$\rm \alpha$ line lies outside the SDSS spectral
range at these redshifts. Therefore, we cannot rule out that these quasars are luminous versions of
the Type $1.9$ Seyferts that show
evidence for a broad H$\rm \alpha$ component but no broad H$\rm \beta$
component \citep{Osterbrock81}.
Second, we selected quasars with low observed
X-ray to \oiii luminosity ratios, $L_{\rm 2-10keV}/L_{\rm
  [OIII]}<2.5$. 
This threshold corresponds to a
two orders of magnitude
suppression of the observed X-ray luminosity,
assuming the \citet{Mulchaey94} relation between \oiii and intrinsic $2$--$10$~keV
flux (taking into account the variance of the relation), which is
consistent with Compton-thick absorption. % see Vignali+10, pg. 53
This is a conservative selection, since the \citet{Mulchaey94} relation was
calibrated for \typeii Seyferts, and \typeii quasars typically have
larger X-ray to \oiii luminosity ratios \citep{Netzer06}.
Third, we made sub-selections of three quasars which show evidence for
extreme obscuration on the basis of different diagnostics: 

\begin{itemize}
\item SDSS~J001111.97+005626.3 \citep[$z=0.409$, $L_{\rm 2-10keV}
= 3.1 \times 10^{42}$~\ergpersec, $L_{\rm [OIII]} = 1.8 \times
10^{42}$~\ergpersec;][]{Reyes08,Jia13} has a flat X-ray spectral slope
at observed-frame $0.3$--$10$~keV
\citep[$\Gamma = 0.6^{+1.17}_{-1.15}$;][]{Jia13}, 
which suggests that the X-ray emission
is rising steeply towards high energies ($>10$~keV). Unlike the other
two quasars, there is no mid-IR spectroscopy available.

\item SDSS~J005621.72+003235.8 \citep[$z = 0.484$, $L_{\rm 2-10keV}
= 8.9 \times 10^{41}$~\ergpersec, $L_{\rm [OIII]} = 6.8 \times
10^{42}$~\ergpersec;][]{Reyes08,Vignali10} has the
deepest $9.7$~$\mu$m silicate (Si) absorption of the sample of \typeii
quasars observed with \spitzer-IRS in
\citet{Zakamska08}. 
Such strong Si features are typically found in Compton-thick AGN
\citep[e.g.,][]{Shi06,Georgantopoulos11a,Goulding12}.

\item SDSS~J115718.35+600345.6 \citep[$z = 0.491$, $L_{\rm 2-10keV}
< 1.5 \times 10^{42}$~\ergpersec, $L_{\rm [OIII]} = 1.6 \times
10^{43}$~\ergpersec;][]{Reyes08,Vignali10} is the most luminous quasar in
the \citet{Vignali10} sample at mid-IR wavelengths, but is undetected by
\chandra \citep{Vignali06}. The extremely low X-ray to mid-IR
luminosity ratio is likely due to Compton-thick absorption \citep{Vignali10}.
The \spitzer-IRS spectrum for this source shows it to be
quasar-dominated at mid-IR wavelengths, but that it also hosts ultraluminous star formation [$\log
(L_{\rm SF}/L_{\rm \odot})=12.3$, \citealt{Zakamska08}]. There
is no evidence for significant Si-absorption; however, $\approx
50\%$ of the best studied Compton-thick AGN do not have significant
Si-absorption \citep[e.g.,][]{Goulding12}.

\end{itemize}

%%%%%%%%%%%%%%%%%%%%%%%%%%%%%%%%%%%%%%%%%%%%%%%%%%%%%%%%%%%%%%%%%%%%%%
\section{NuSTAR and Multiwavelength Data}
\label{data}
%%%%%%%%%%%%%%%%%%%%%%%%%%%%%%%%%%%%%%%%%%%%%%%%%%%%%%%%%%%%%%%%%%%%%%

In our analysis of the three \typeii quasars, we used \nustar
observations in conjunction with lower energy X-ray observations from \chandra
and \xmm, and near-UV to mid-IR data primarily from large-area public
surveys. Hereafter we refer to the quasars using abbreviated
SDSS object names.

\subsection{NuSTAR Observations}
\label{nustar_data}

\nustar consists of two co-aligned X-ray telescopes
(focal length $= 10.14$~m) which use grazing incidence optics to focus
hard X-rays ($3$--$79$~keV) onto two focal plane modules \citep[FPMs A and B;][]{Harrison13}.
Each FPM provides a $\approx 12\arcmin \times 12\arcmin$ field of view at 10
keV, and a pixel size of $2.46$\arcsec.
The \nustar PSF is characterized by a full-width half maximum (FWHM) of $18$\arcsec, and a half-power diameter of $58$\arcsec. The
astrometric accuracy for bright X-ray sources is $\pm
8$\arcsec\ \citep[$90\%$ confidence;][]{Harrison13}.

The \typeii quasars, \sdssa, \sdssb and \sdssc, were observed by \nustar with nominal exposure times
of $19.6$~ks, $23.5$~ks and $23.3$~ks, respectively. Details of the observations, including net exposure
times, are
provided in Table \ref{obs_table}.
\begin{table*}[]
\centering
\caption{X-ray Observation Log}
%\centering
\begin{tabular}{lcccccccc} \hline\hline \noalign{\smallskip}
\multicolumn{2}{l}{} & \multicolumn{3}{c}{\nustar} &
\multicolumn{4}{c}{Lower Energy X-ray Observations} \\
\noalign{\smallskip}
\cmidrule(rl){3-5} \cmidrule(rl){6-9}
\noalign{\smallskip}
\multicolumn{1}{c}{Object Name} & $z$ & Observation ID & UT Date & Exposure & Observatory &
Observation ID & UT Date & Exposure \\
\multicolumn{1}{c}{(1)} & (2) & (3) & (4) & (5) & (6) & (7) & (8) &
(9) \\
\noalign{\smallskip} \hline \noalign{\smallskip}
SDSS J001111.97+005626.3 & 0.409 & 60001065002 & 2013 Jan 27 & 18.3 ks & \xmm & 0403760301
& 2006 Jul 10 & 25.7 ks \\
SDSS J005621.72+003235.8 & 0.484 & 60001061002 & 2013 Jan 27 & 21.9 ks
& \chandra & 7746 & 2008 Feb 08 & 9.91 ks \\
SDSS J115718.35+600345.6 & 0.491 & 60001071002 & 2012 Oct 28 & 21.7 ks
& \chandra & 5698 & 2005 Jun 03 & 6.97 ks \\
\noalign{\smallskip} \hline \noalign{\smallskip}
\end{tabular}
\begin{minipage}[l]{0.915\textwidth}
\footnotesize
NOTE. -- (1): Full SDSS object name. 
(2): Redshift. 
(3) and (4): \nustar observation ID and start date.
(5): Net on-axis \nustar exposure time. 
This value applies to both FPMA and FPMB.
(6) Lower energy X-ray observatory data used (\chandra or \xmm).
(7), (8) and (9): \chandra or \xmm observation ID, observation start date, and net
on-axis exposure time, corrected for flaring and bad events.  
\end{minipage}
\label{obs_table}
\end{table*}
We processed the data using the {\it NuSTAR} Data Analysis
Software (NuSTARDAS) v. 1.3.0. Calibrated and cleaned event files were produced using the
{\sc nupipeline} script and the {\it NuSTAR} CALDB 20131007 release with
the standard filter flags.

\subsubsection{Photometry and Source Detection}
\label{photometry}

To characterize the high energy X-ray emission and determine whether
sources are detected, we performed photometry in the observed-frame $3$--$24$~keV, $3$--$8$~keV, and
$8$--$24$~keV bands for both of the \nustar FPMs following
\citet{Alexander13}. We avoided using photons above $24$~keV, where
the drop in effective area and the prominent
background features (see Figure 2 and 10 of
\citealt{Harrison13}, respectively) 
hinder the analysis of faint X-ray
sources such as \typeii quasars. We split the \nustar
event files into individual band images using {\sc
  dmcopy}, part of the {\it Chandra} Interactive Analysis Observations software ({\sc
  CIAO}, v4.4;
\citealt{Fruscione06}).\footnote{http://cxc.harvard.edu/ciao/index.html} 
We extracted the gross source counts ($S$) from a $45$\arcsec\ radius aperture centered on the SDSS
position. For a source at the \nustar aim point, and for the
  energy range ($3$--$24$~keV) and 
  spectral slopes ($\Gamma=0.6$--$1.8$) used in this study, this aperture
encloses $\approx65\%$ of the full PSF energy.
We extracted the background counts ($B$) from an annulus with an
inner radius $90$\arcsec\ from the source and an outer radius
$150$\arcsec\ from the source, which allowed the local
background to be sampled while minimising
contamination from the source. To obtain
the background counts in the source extraction region ($B_{\rm src}$),
we multiplied $B$ by the area scaling factor between the source and
background regions ($A_S/A_B$). Net
source counts were calculated as $S-B_{\rm src}$, and corresponding $68.3\%$
confidence level uncertainties were taken as $\sqrt{S+B(A_S/A_B)^2}$.
For non detections, we calculated $99.7\%$
confidence level upper limits using the Bayesian method of \citet{Kraft91}. The \nustar photometry is given in Table \ref{xray_photometry}.
\begin{table*}[]
\centering
\caption{X-ray Photometry}
%\centering
\begin{tabular}{lcccccccccc} \hline\hline \noalign{\smallskip}
\multicolumn{1}{c}{Object Name} & \multicolumn{3}{c}{Net Counts (FPMA)} &
\multicolumn{3}{c}{Net Counts (FPMB)} & \multicolumn{3}{c}{Flux
  (\nustar)} & \multicolumn{1}{c}{Flux (other)} \\
\multicolumn{1}{c}{(1)} & \multicolumn{3}{c}{(2)} &
\multicolumn{3}{c}{(3)} & \multicolumn{3}{c}{(4)} & \multicolumn{1}{c}{(5)} \\
\noalign{\smallskip}
\cmidrule(rl){2-4} \cmidrule(rl){5-7} \cmidrule(rl){8-10} \cmidrule(rl){11-11}
\noalign{\smallskip}
\multicolumn{1}{c}{} & $3$--$24$ & $3$--$8$ &
$8$--$24$ & $3$--$24$ & $3$--$8$ & $8$--$24$ & $3$--$24$ & $3$--$8$ &
$8$--$24$ & $3$--$8$ \\
%\multicolumn{1}{c}{and FPM} & (ks) & \\
%\multicolumn{1}{c}{(1)} & (2) & (3) & (4) & (5) & (6) & (7) & (8) & (9) \\
\noalign{\smallskip} \hline \noalign{\smallskip}
0011+0056 & $17.5  \pm 7.7$ & $<16.3$ & $16.8 \pm 6.4$ & $<24.7$ & $<18.3$ & $<19.7$ & $0.99$ & $<0.74$ & $1.32$ & $0.18$  \\
0056+0032 & $<19.1$ & $<10.9$ & $<20.9$ & $<23.5$ & $<17.8$ & $<19.6$ & $<1.14$ & $<0.52$ & $<1.58$ & $<0.16$ \\
1157+6003 & $<31.4$ & $<16.6$ & $<29.1$ & $<23.3$ & $<20.0$ & $<17.7$ & $<1.35$ & $<0.63$ & $<1.68$ & $<0.22$  \\
\noalign{\smallskip} \hline \noalign{\smallskip}
\end{tabular}
\begin{minipage}[l]{0.82\textwidth}
\footnotesize
NOTE. -- (1): Abbreviated SDSS object name. The `SDSS J' prefix and
all RA and Dec digits after the first four have been truncated. (2)
and (3): Net source counts in the observed-frame $3$--$24$,
$3$--$8$ and $8$--$24$~keV bands for FPMA and FPMB,
respectively. $68.3\%$ confidence level
uncertainties, and $99.7\%$ confidence level upper limits
are given. (4): Aperture-corrected \nustar flux in units of
$10^{-13}$~\fluxunit (for a power-law model with $\Gamma=1.8$), in the
observed-frame $3$--$24$,
$3$--$8$ and $8$--$24$~keV bands. For \sdssa the fluxes are for FPMA
only, while for \sdssb and \sdssc the fluxes are averaged over FPMA
and FPMB. (5): Aperture-corrected observed-frame $3$--$8$~keV flux in units of
$10^{-13}$~\fluxunit (for a power-law model with $\Gamma=1.8$), as measured using
lower-energy X-ray data. \xmm data have been used for \sdssa, and
\chandra data have been used for \sdssb and \sdssc. $99.7\%$ confidence level upper limits
are given. 
\end{minipage}
\label{xray_photometry}
\end{table*}

\label{source_detection}

To test whether the quasars are detected in the individual \nustar band
images, we looked for significant source signals at their SDSS positions.
We assumed binomial statistics and calculated
false probabilities, or `no-source' probabilities ($P$), using the following equation:
%\begin{eqnarray}
%P_B(S) = \frac{ \int_0^B e^{-t} t^{S-1} {\rm d}t }{ \int_0^\infty
%  e^{-t} t^{S-1} {\rm d}t } .
%\label{eq_fprob}
%\end{eqnarray}
\begin{eqnarray}
P(x\ge S) =
\sum\limits_{x=S}^{T}\frac{T!}{x!(T-x)!}p^{x}\left(1-p\right)^{T-x} ,
\label{eq_fprob}
\end{eqnarray}
where $T=S+B$ and $p = 1/(1+B/B_{\rm src})$. 
$P$ is the probability that, assuming there is no source at
the SDSS position, the measured gross counts in the source aperture ($S$)
are purely due to a background fluctuation \citep[][]{Weisskopf07}.

\begin{figure}
\centering
\includegraphics[width=0.47\textwidth]{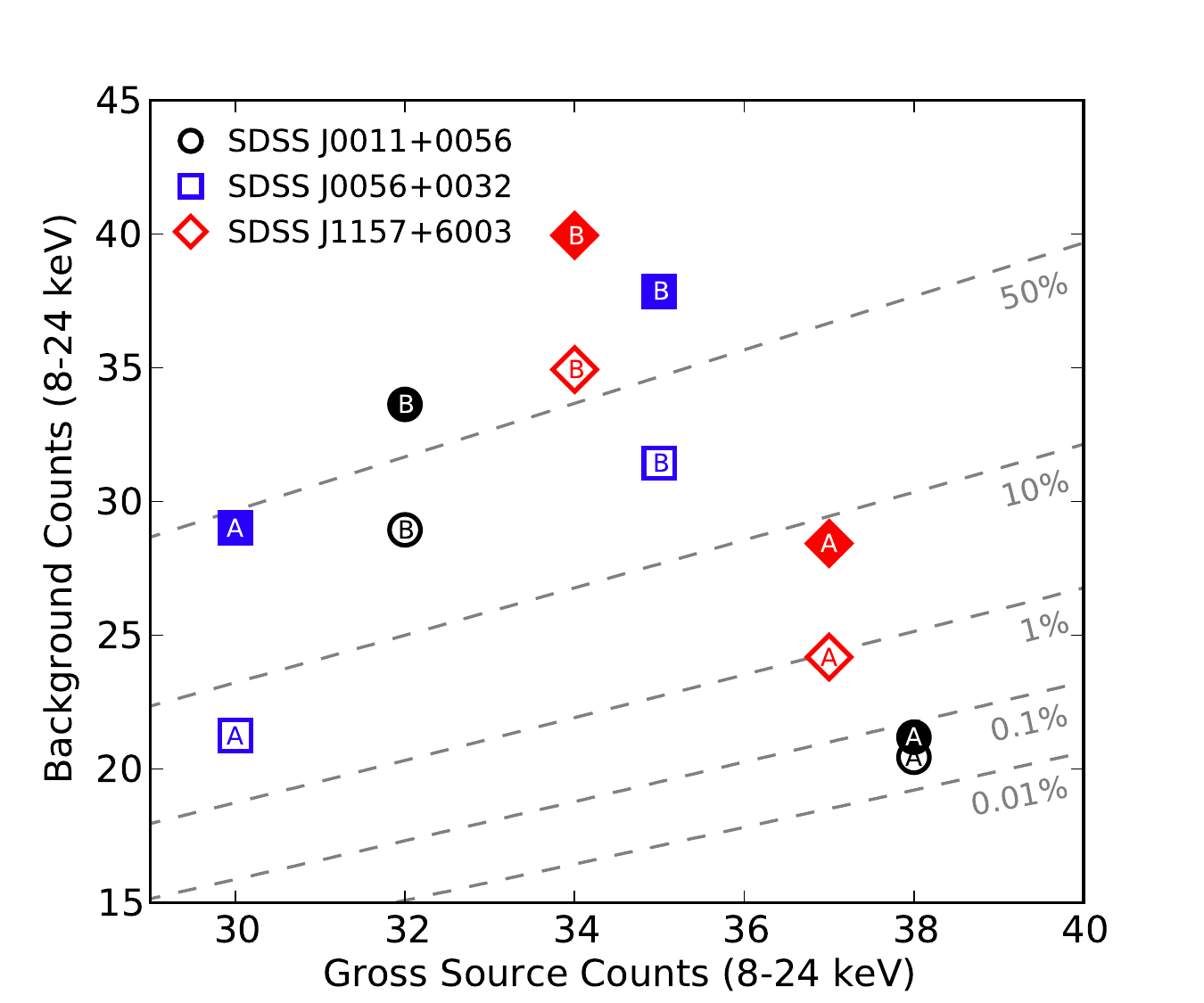}
\caption{Gross source counts ($S$) versus scaled background counts
  ($B_{\rm src}$) at
  observed-frame $8$--$24$~keV for \sdssa, \sdssb and \sdssc (circles, squares
 and diamonds, respectively). Background counts were measured using two approaches:
  direct measurement from the \nustar images (filled symbols), and from model background
  maps (empty symbols). The A and B labels correspond to FPMA and
  FPMB, respectively. The dashed lines indicate Poisson no-source
  probabilities. There is one significant detection: SDSS~J0011+0056
  is detected with FPMA.}
\label{false_probs}
\end{figure}

\begin{figure}
\centering
\includegraphics[width=0.47\textwidth]{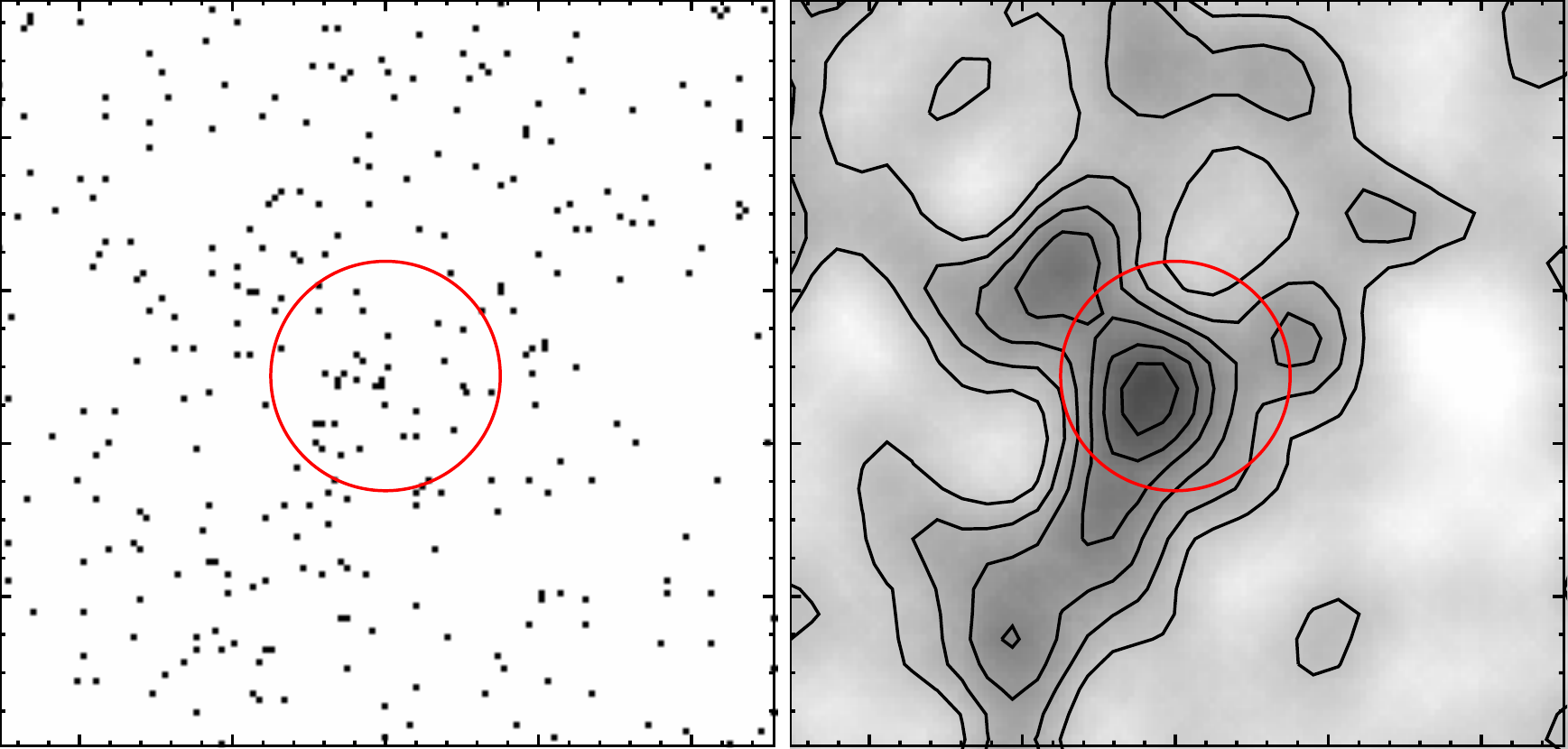}
\caption{{\it NuSTAR} FPMA $8$--$24$~keV image centered on the
  SDSS position of 
  SDSS J0011+0056. Left panel: Unsmoothed image. Right panel: Image
  smoothed with a Gaussian of radius $14$ pixels ($34.5$\arcsec), and with
  overlaid contours of constant pixel values. The smoothing and contours
  are for display purposes only. A $45$\arcsec\ radius aperture is
  shown (red circle), centered on the SDSS position. The horizontal
  and vertical axes are right ascension
  (RA) and declination (Dec), respectively. The major ticks indicate
  1 arcmin offsets.}
\label{SDSS0011_detection}
\end{figure}

Given that the three \typeii quasars are faint at $3$--$8$~keV (see
Table \ref{xray_photometry} for \chandra and \xmm fluxes and upper limits), and likely have flat X-ray spectra with
emission rising steeply to higher energies,
\nustar is most likely to detect the sources above
$8$~keV (observed-frame).
At these energies \chandra and \xmm have little to no sensitivity. 
In Figure \ref{false_probs}, we show the $S$ and $B_{\rm src}$ values
measured with \nustar 
for the $8$--$24$~keV band (filled symbols), and the no-source probabilities to which
they correspond (dashed lines).\footnote{We avoid overplotting the
  errors for individual $S$ and $B_{\rm src}$ measurements, since these are not used in
   the calculation of no-source probabilities.}
For the purposes of this figure, Poisson statistics
have been assumed; for our sources, $B$ is large and the Poisson
integral thus provides a good approximation of Equation \ref{eq_fprob} \citep[][]{Weisskopf07}.
Taking binomial no-source probabilities greater than $1 \%$ to indicate
non detections, neither SDSS~J0056+0032 nor SDSS~J1157+6003 are detected in either FPM.
SDSS~J0011+0056, on the other hand, is detected in FPMA with a binomial
no-source probability of $0.093\%$.\footnote{We note that, in
  this case, using a $50$\arcsec\ (as opposed to $45$\arcsec) source aperture results in a lower
  no-source probability of $0.049\%$.} The \nustar image corresponding to this detection is shown
in Figure \ref{SDSS0011_detection}. 
The source is not detected in FPMB,
which has higher background
noise relative to FPMA for this observation; indeed the net source
counts for FPMA are consistent with the upper
limit for FPMB (see Table
\ref{xray_photometry}). SDSS~J0011+0056 is also weakly detected in the
$3$--$24$~keV band for FPMA, with a binomial no-source probability of $0.58\%$. Aside from this, none of the quasars are detected in the
$3$--$8$~keV and $3$--$24$~keV bands. 

The no-source probability is sensitive to the background region
sampled. To partially address this we also measured the background from model background
maps produced using {\sc nuskybgd} (Wik et al., in prep.), summing
counts within the $45$\arcsec\ radius source aperture.
These measurements are shown as empty symbols in Figure
\ref{false_probs}. 
SDSS~J0011+0056 is still detected in FPMA using this
approach, with a no-source probability of $0.033\%$
at $8$--$24$~keV.

\subsubsection{Flux Calculation}
\label{flux_calc}

For each \nustar energy band we determined the conversion factor between net
count rate and source flux using {\sc XSPEC}
v12.8.1j \citep{Arnaud96}, taking into account the Response Matrix
File (RMF) and Ancillary Response File (ARF) for each FPM.
We assumed a power-law model with $\Gamma = 1.8$,
consistent with that found for AGN at observed-frame $3$--$24$~keV
\citep{Alexander13}. 
%We note that Galactic extinction is negligible
%at these high X-ray energies.
We corrected fluxes to the $100\%$ encircled-energy fraction of the PSF.
The \nustar fluxes are given in Table \ref{xray_photometry}.

For the \nustar-detected quasar, \sdssa, we measure an observed-frame $8$--$24$~keV flux of $1.32 \times
10^{-13}$~\fluxunit. 
This value is consistent with
extrapolations from the \xmm
$0.5$--$10$~keV count rate given the photon
index constraints of \citet{Jia13}, $\Gamma = 0.6^{+1.17}_{-1.15}$,
and assuming a simple unabsorbed power-law model. 
Additionally, as we later show in
Section \ref{results_indirect}, our X-ray flux measurement
for \sdssa is consistent with that expected from its \sixum
luminosity, which is assumed to result from the reprocessing of AGN emission by obscuring dust.

\subsection{Lower Energy X-ray Data}
\label{low_data}

For \sdssa we used the archival {\it XMM-Newton} EPIC observation, first
published in \citet{Jia13}.
We analyzed the Pipeline Processing System (PPS) data products for this observation 
using the \emph{Science Analysis
  Software}\footnote{http://xmm.esa.int/sas/} (SAS v.12.0.1). 
The MOS1 and MOS2 data were coadded with the SAS task
{\sc epicspeccombine}. The PN data were excluded, since \sdssa is close
to a chip gap.
The source counts were extracted from a $15$\arcsec\ radius aperture
and the
background counts were extracted using an 
$80$\arcsec\ radius source-free aperture, selected to sample the
local background while avoiding chip gaps and nearby serendipitous sources.
We used {\sc XSPEC} to convert from count rate to flux, assuming a
power-law model with $\Gamma = 1.8$ and using the \xmm RMF and
ARF.
Throughout this work, we neglect the cross-calibration constants
between MOS and \nustar as the current best estimates are
  $\sim7\pm5\%$ (Madsen et al., in prep.), and a change on this scale does not affect our results.

For \sdssb and \sdssc we used the archival {\it Chandra} observations,
first published in
\citet{Vignali06,Vignali10}. 
We reprocessed the data using {\sc
  chandra\_repro},\footnote{http://cxc.harvard.edu/ciao/ahelp/chandra\_repro.html}
a CIAO pipeline, to create event files. 
The source counts were extracted from a $3$\arcsec\ radius aperture,
and the background counts were extracted from
an annulus with an inner radius $10$\arcsec\ from
the source and an outer radius $30$\arcsec\ from the source. 
As \sdssb and \sdssc are non detections at observed-frame $3$--$8$~keV, we calculated $99.7\%$ confidence level 
upper limits for the source counts using the Bayesian method of \citet{Kraft91}.
To calculate fluxes, we converted from \chandra count rates with the HEASARC tool
WebPIMMs\footnote{http://heasarc.gsfc.nasa.gov/Tools/w3pimms.html}
(v4.6b) assuming a power-law model with $\Gamma=1.8$, and corrected to the $100\%$ encircled-energy fraction of the PSF.

As the \typeii quasars are faint at X-ray wavelengths, we are unable to fit
the spectra accurately. For instance, \sdssa is detected with \xmm, 
but using the combined MOS1+MOS2 data we only extract $5.6$ and $20.6$ 
net source counts in the observed-frame $0.5$--$3$~keV and $3$--$8$~keV bands, respectively.
We list the \chandra and \xmm $3$--$8$~keV fluxes and upper limits in
Table \ref{xray_photometry}.

\subsection{Near-UV to Mid-IR Data and SED Decomposition}
\label{UVMIR_data}

To investigate the multiwavelength properties of the three \typeii quasars,
in particular the mid-IR emission from the AGN, we
collected photometric data at $0.3$--$30$~$\mu$m (i.e., at near-UV through mid-IR
wavelengths). We used imaging data from public large-area surveys,
primarily
the Sloan Digital Sky Survey \citep[SDSS;][]{York00}, 
the UKIRT Infrared Deep Sky Survey \citep[UKIDSS;][]{Lawrence07},
and the {\it WISE} all-sky survey \citep[][]{Wright10}. Additionally, for
\sdssb and \sdssc, we used {\it Spitzer}
photometry from the {\it Spitzer} Enhanced Imaging Products Source List.\footnote{http://irsa.ipac.caltech.edu/data/SPITZER/Enhanced/Imaging/}
The photometric dataset, not corrected for Galactic extinction, is
provided in Table~\ref{UVMIR_table}.
We note that since the observations are not contemporaneous, AGN
variability may affect the SED analysis at longer
wavelengths, where the AGN is bright with respect to the host galaxy.
\begin{table}
\centering
\caption{Near-Ultraviolet to Mid-Infrared Source Properties}
%\centering
\begin{tabular}{lccc} \hline\hline \noalign{\smallskip}
Object Name$^{a}$	&	0011+0056	&	0056+0032	&	1157+6003	\\ 
\noalign{\smallskip} \hline \noalign{\smallskip}
% NB: the SDSS central wavelengths are those I used for the SED
% fitting.

$u$ (0.355~$\mu$m)$^{b}$	&	$23.51\pm0.99$	&	$23.25\pm0.71$	&	$20.53\pm0.06$	\\
$g$ (0.468~$\mu$m)$^{b}$	&	$21.50\pm0.05$	&	$21.60\pm0.069$	&	$20.10\pm0.01$	\\
$r$ (0.616~$\mu$m)$^{b}$	&	$20.26\pm0.05$	&	$20.72\pm0.05$	&	$19.61\pm0.02$	\\
$i$ (0.748~$\mu$m)$^{b}$	&	$19.60\pm0.04$	&	$19.82\pm0.04$	&	$18.90\pm0.01$	\\
$z$ (0.892~$\mu$m)$^{b}$	&	$19.25\pm0.09$	& $19.81\pm0.12$	&	$19.02\pm0.05$	\\

$Y$ (1.03~$\mu$m)$^{c}$	&	$18.25\pm0.04$	&	$-$	&	$-$	\\
$J$ (1.25~$\mu$m)$^{c}$	&	$17.70\pm0.03$	&	$18.42\pm0.07$	&	$-$	\\
$H$ (1.63~$\mu$m)$^{c}$	&	$16.81\pm0.04$	&	$17.31\pm0.07$	&	$-$	\\
$K$ (2.20~$\mu$m)$^{c}$	&	$-$	&	$16.64\pm0.05$	&   $-$	\\

{\it WISE} (3.4~$\mu$m)$^{d}$	&	$14.94\pm0.04$	&	$15.53\pm0.05$	&	$12.78\pm0.02$	\\
{\it WISE} (4.6~$\mu$m)$^{d}$	&	$14.45\pm0.07$	&	$14.51\pm0.08$	&	$11.24\pm0.02$	\\
{\it WISE} (12~$\mu$m)$^{d}$	&	$10.62\pm0.09$	&	$9.77\pm0.05$	&	$8.0\pm0.02$	\\
{\it WISE} (22~$\mu$m)$^{d}$	&	$-$	&	$6.55\pm0.07$ &	$5.37\pm0.03$	\\

{\it Spitzer} (3.6~$\mu$m)$^{e}$	&	$-$	&	$0.173\pm0.003$	&	$2.860\pm0.009$	\\
{\it Spitzer} (4.5~$\mu$m)$^{e}$	&	$-$	&	$0.220\pm0.003$	&	$4.511\pm0.010$	\\
{\it Spitzer} (5.8~$\mu$m)$^{e}$	&	$-$	&	$0.591\pm0.009$	&	$8.215\pm0.017$	\\
{\it Spitzer} (8.0~$\mu$m)$^{e}$	&	$-$	&	$2.474\pm0.016$	&	$13.165\pm0.022$	\\
{\it Spitzer} (24~$\mu$m)$^{f}$	&	$-$	& $18.088\pm0.058$  & $57.318\pm0.062$	\\

{\it IRAS} (60~$\mu$m)$^{g}$	&	$-$	& $-$  & $260.0\pm46.0$	\\

% OLD NUMBERS (in original units):
%$u$ (0.355~$\mu$m)$^{b}$	&	$23.51\pm0.99$	&	$23.25\pm0.71$	&	$20.53\pm0.06$	\\
%$g$ (0.468~$\mu$m)$^{b}$	&	$21.50\pm0.05$	&	$21.60\pm0.069$	&	$20.10\pm0.01$	\\
%$r$ (0.616~$\mu$m)$^{b}$	&	$20.26\pm0.05$	&	$20.72\pm0.05$	&	$19.61\pm0.02$	\\
%$i$ (0.748~$\mu$m)$^{b}$	&	$19.60\pm0.04$	&	$19.82\pm0.04$	&	$18.90\pm0.01$	\\
%$z$ (0.892~$\mu$m)$^{b}$	&	$19.25\pm0.09$	&	$19.81\pm0.12$	&	$19.02\pm0.05$	\\
%$Y$ (1.03~$\mu$m)$^{c}$	&	$18.25\pm0.04$	&	$-$	&	$-$	\\
%$J$ (1.25~$\mu$m)$^{c}$	&	$17.70\pm0.03$	&	$18.42\pm0.07$	&	$-$	\\
%$H$ (1.63~$\mu$m)$^{c}$	&	$16.81\pm0.04$	&	$17.31\pm0.07$	&	$-$	\\
%$K$ (2.20~$\mu$m)$^{c}$	&	$-$	&	$16.64\pm0.05$	&	$-$	\\
%{\it WISE} (3.4~$\mu$m)$^{d}$	&	$14.94\pm0.04$	&	$15.53\pm0.05$	&	$12.78\pm0.02$	\\
%{\it WISE} (4.6~$\mu$m)$^{d}$	&	$14.45\pm0.07$	&	$14.51\pm0.08$	&	$11.24\pm0.02$	\\
%{\it WISE} (12~$\mu$m)$^{d}$	&	$10.62\pm0.09$	&	$9.77\pm0.05$	&	$8.0\pm0.02$	\\
%{\it WISE} (22~$\mu$m)$^{d}$	&	$-$	&	$6.55\pm0.07$	&	$5.37\pm0.03$	\\
%{\it Spitzer} (3.6~$\mu$m)$^{e}$	&	$-$	&	$0.173\pm0.003$	&	$2.860\pm0.009$	\\
%{\it Spitzer} (4.5~$\mu$m)$^{e}$	&	$-$	&	$0.220\pm0.003$	&	$4.511\pm0.010$	\\
%{\it Spitzer} (5.8~$\mu$m)$^{e}$	&	$-$	&	$0.591\pm0.009$	&	$8.215\pm0.017$	\\
%{\it Spitzer} (8.0~$\mu$m)$^{e}$	&	$-$	&	$2.474\pm0.016$	&	$13.165\pm0.022$	\\
%{\it Spitzer} (24~$\mu$m)$^{f}$	&	$-$	& $18.088\pm0.058$	&	$57.318\pm0.062$	\\
\noalign{\smallskip}
\noalign{\smallskip}
\noalign{\smallskip}
$\hat{a}$$^{h}$	&	$0.590\pm0.029$	&	$0.946\pm0.003$	&	$0.977\pm0.001$	\\
$L_{\rm 6\mu m}$$^{h}$	&	$1.14\pm0.15$	&	$15.19\pm0.60$	&	$51.44\pm1.12$	\\

\noalign{\smallskip} \hline \noalign{\smallskip}
\end{tabular}
\begin{minipage}[l]{0.475\textwidth}
\footnotesize
NOTE. -- 
$^{a}$ Abbreviated SDSS object name;
$^{b}$ SDSS DR7 Fiber magnitudes in the AB sinh system; 
$^{c}$ UKIDSS DR9 2.8\arcsec\ diameter aperture magnitudes in the Vega system;
$^{d}$ {\it WISE} profile-fit magnitudes in the Vega system; 
$^{e}$ {\it Spitzer} 3.8\arcsec\ diameter aperture flux densities in
units of mJy; 
$^{f}$ {\it Spitzer} PSF-fit flux densities in units of mJy; 
$^{g}$ {\it IRAS} flux density in units of mJy \citep{Zakamska04}. This data
point was not used in the SED modeling; 
$^{h}$ best-fit parameters (corrected for dust reddening) from the SED decomposition
described in Section \ref{UVMIR_data}:
$\hat{a}$ is the fractional contribution of the AGN to the
$0.1$--$30$~$\mu$m emission; $L_{\rm 6\mu m}$ is the rest-frame \sixum
luminosity (${\rm \nu} L_{\rm \nu}$) of the
AGN in units of $10^{44}$~erg~s$^{-1}$. The
uncertainties are standard deviations, derived from the Monte Carlo
re-sampling of the photometric data.
\end{minipage}
\label{UVMIR_table}
\end{table}

\begin{figure*}
\centering
\includegraphics[width=1\textwidth]{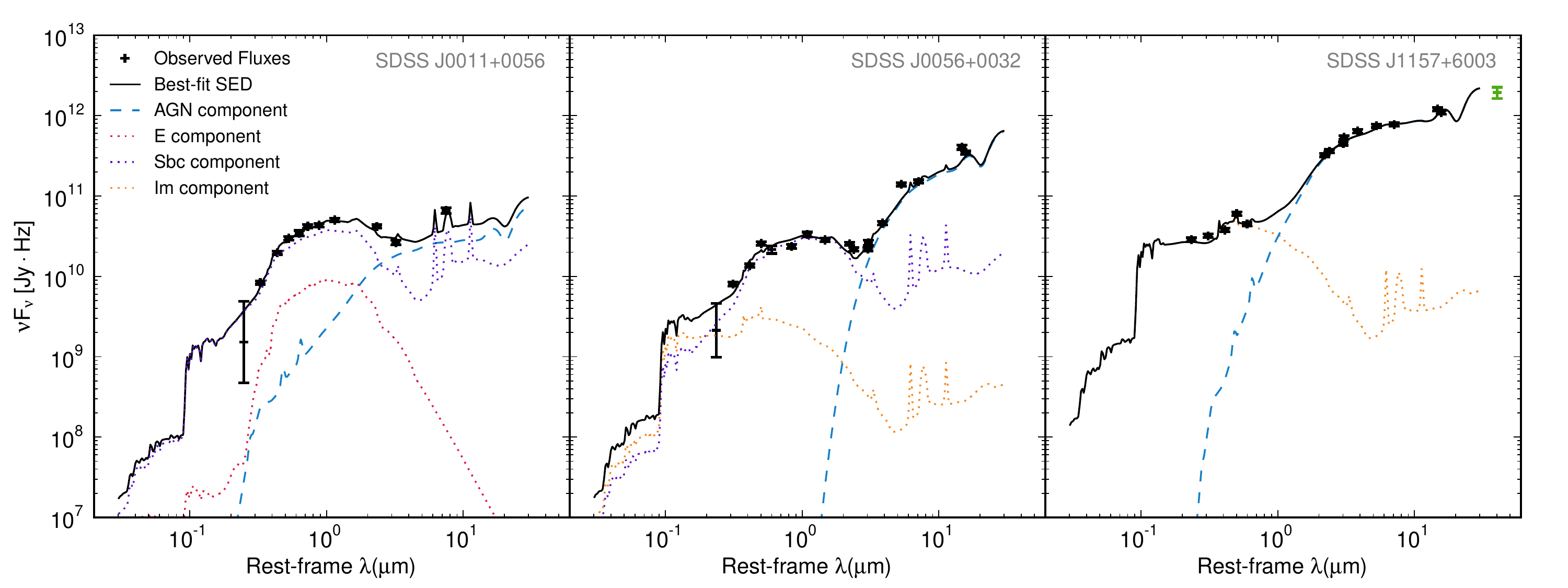}
\caption{Near-UV to mid-IR SEDs for the three \typeii quasars. The best-fitting
  model solutions (black line) were achieved using the AGN (blue
  dashed line) and galaxy (elliptical: red dotted line, spiral: purple dotted line,
  irregular: orange dotted line) templates of
  \citet{Assef10}. The
  photometric data (black data points) and best-fitting
  parameters are given in Table \ref{UVMIR_table}. The {\it IRAS}
  60~$\mu$m flux for \sdssc (green data point) was not used in the SED decomposition.}
\label{SED_figure}
\end{figure*}

We used the near-UV through mid-IR photometric data to
produce broad-band spectral energy distributions (SEDs) for our sample.
We modeled the SEDs using the \citet{Assef10} 0.03--30~$\mu$m empirical low-resolution AGN and
galaxy templates.
Each SED was modeled as a best-fit
combination of an elliptical, a spiral and an irregular
galaxy component, plus an AGN.
We refer the reader to \citet{Assef08,Assef10,Assef13} for further details.
In Fig.~\ref{SED_figure} we present the SEDs and best-fitting
model solutions. 
For \sdssc we also show the {\it IRAS} 60~$\mu$m
  flux measured by \citet[][green data point in Fig.~\ref{SED_figure}]{Zakamska04}, which lies beyond the wavelength range of the galaxy
  templates and was therefore excluded from the SED modeling. The data
  point is consistent with a simple extrapolation of the best-fitting
  model from shorter wavelengths. \citet{Zakamska04} also detect
  \sdssb at 60~$\mu$m, but at a low significance level ($80\%$
  confidence).
In Table~\ref{UVMIR_table} we provide the best-fitting parameters $\hat{a}$
(the fractional contribution from the AGN component to the
$0.1$--$30$~$\mu$m emission after correction for dust reddening; \citealt{Assef10}) and $L_{\rm 6\mu m}$ (the luminosity of the AGN
component at rest-frame \sixum after correction for dust reddening; ${\rm \nu} L_{\rm \nu}$). 
The uncertainties on $\hat{a}$ and $L_{\rm 6\mu m}$ are standard
deviations, derived from the Monte Carlo re-sampling of the
data according to the photometric uncertainties. Both
  parameters are well constrained.\footnote{Constraining
    $\hat{a}$ and $L_{\rm 6\mu m}$ is the primary purpose of our SED
    analysis. We do not read deeply into the host-galaxy properties of the best fitting solutions.}
Since our SED modeling uses a single AGN template, it does not account
for the fact that AGN show a range of heated dust
emission relative to the bolometric emission of the accretion
disk. For instance, assuming the distribution of quasar covering factors found
by \citet{Roseboom13} would introduce an additional uncertainty to the \sixum
luminosities of $\approx \pm 0.5 L_{\rm 6\mu m}$.
Our three \typeii quasars have high $\hat{a}$ values, which indicates that they are AGN-dominated at
$0.1$--$30$~$\mu$m. For \sdssb and \sdssc this is in agreement with
the {\it Spitzer}-IRS spectroscopy of \citet{Zakamska08}, which shows the sources to be
AGN-dominated at mid-IR wavelengths (for \sdssa there is no mid-IR
spectroscopy available). 

%%%%%%%%%%%%%%%%%%%%%%%%%%%%%%%%%%%%%%%%%%%%%%%%%%%%%%%%%%%%%%%%%%%%%%
\section{Results}
\label{Results}
%%%%%%%%%%%%%%%%%%%%%%%%%%%%%%%%%%%%%%%%%%%%%%%%%%%%%%%%%%%%%%%%%%%%%%

The three \typeii quasars in this work bear the signatures of heavily
obscured, Compton-thick AGN based on multi-wavelength diagnostics
\citep[see Section 2 of this
work;][]{Zakamska08,Vignali10,Jia13}. Here we present the results of
our analysis, which is aimed at assessing the
prevalence of extreme absorption in these systems.

X-rays provide a direct measure of AGN emission that has been subject
to circumnuclear absorption. As such, the characterisation of X-ray spectra is necessary to obtain
reliable estimates of absorbing column densities ($N_{\rm
  H}$).\footnote{All $N_{\rm H}$ values in this Section are
  line-of-sight column densities unless otherwise stated. In the {\sc
  MYTorus} model, $N_{\rm H}$ is related to the equatorial column density
  ($N_{\rm H,eq})$ via Equation 1 in \citet{Murphy09}.}
For \sdssa we detect X-rays over the observed-frame $3$--$24$~keV energy range,
and for \sdssb and \sdssc we place upper limits on the
$3$--$24$~keV emission (see Table \ref{xray_photometry}). As the quasars are at
best weak detections at $3$--$24$~keV, detailed
modeling of their X-ray spectra is unfeasible.
For \sdssa we characterize the observed-frame $3$--$24$~keV X-ray spectrum using the
ratio of hard ($8$--$24$~keV) to soft ($3$--$8$~keV) emission, which
provides a direct absorption constraint (see
Section \ref{results_direct}). For the remaining
two quasars we are limited to indirect absorption constraints
from the comparison of the observed X-ray emission with the
intrinsic X-ray emission implied by infrared measurements (see
Section \ref{results_indirect}).

\subsection{Direct (X-ray) Absorption Constraints}
\label{results_direct}

%%%%%%%%%%%%%%
% X-ray properties %
%%%%%%%%%%%%%%

\begin{figure}
\centering
\includegraphics[width=0.47\textwidth]{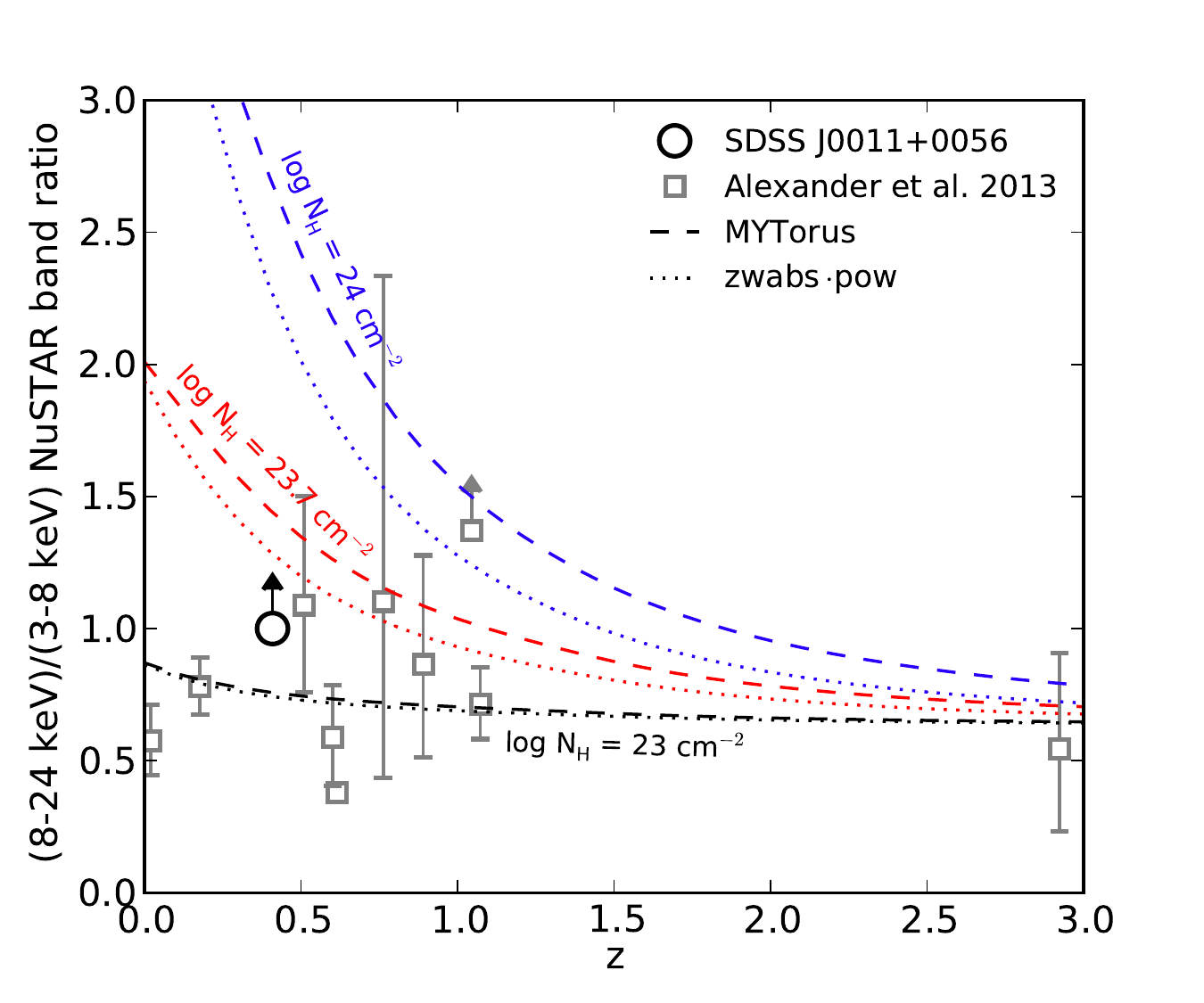}
\caption{\nustar X-ray band ratio ($8$--$24$~keV to $3$--$8$~keV counts ratio)
  against redshift for \sdssa (black circle), and the \nustar-detected
  sources in \citet{Alexander13} (grey squares). The dashed and dotted lines show band ratio
  predictions from {\sc MYTorus} and simple {\sc
    zwabs$\cdot$pow} models respectively, for a
  variety of column densities, and assuming a spectral slope of
  $\Gamma = 1.8$. Varying $\theta_{\rm obs}$ makes a negligible
  difference to the {\sc MYTorus} tracks.
  Based on the $99.7\%$ lower limit for the \nustar band ratio, \sdssa is
  consistent with being heavily obscured.}
\label{bandratio_z}
\end{figure}

\begin{figure}
\centering
\includegraphics[width=0.47\textwidth]{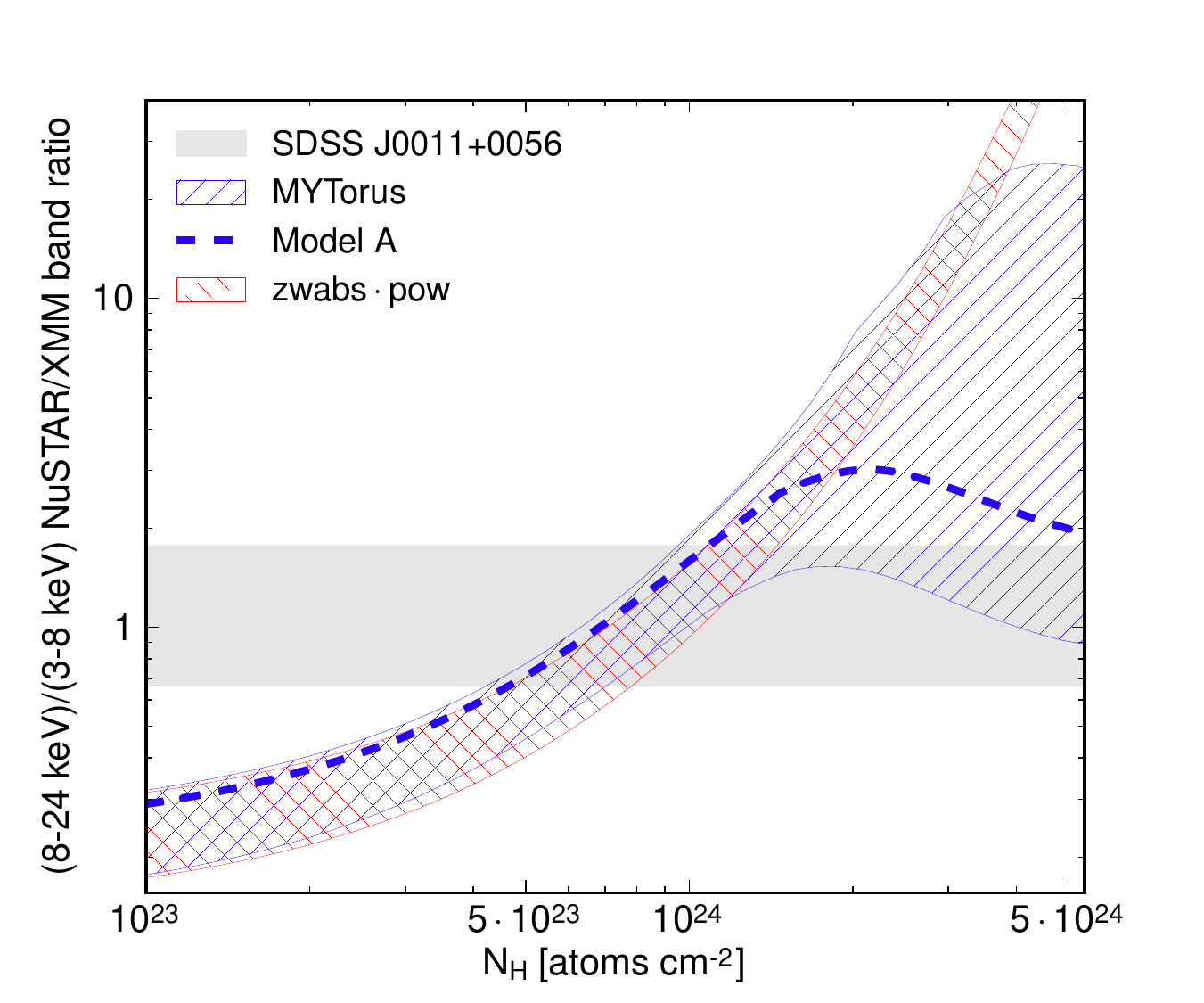}
\caption{\nustar/\xmm X-ray band ratio (\nustar $8$--$24$~keV to \xmm $3$--$8$~keV count-rate ratio) against
  line-of-sight X-ray absorbing column density ($N_{\rm H}$). The
  grey shaded area shows the $68.3\%$ confidence level region for the observed band ratio of
  \sdssa. The hashed regions show the range of band ratios
  predicted with {\sc MYTORUS} (blue) and a simple {\sc
    zwabs$\cdot$pow} model (red) for $z = 0.409$, and
  for a range of intrinsic photon indices ($1.7<\Gamma<2.3$). The {\sc
    MYTorus} region was computed for a range of
  inclination angles ($65^{\circ}<\theta_{\rm obs}<90^{\circ}$).
According to these models, \sdssa is absorbed by $N_{\rm H} \gtrsim 5
\times 10^{23}$~\nhunit. We also show band ratio predictions for a
specific {\sc MYTorus} model with $\Gamma=1.8$ and $\theta_{\rm
  obs}=70^{\circ}$ ({\it Model A}; dashed blue line); on the basis
of {\it Model A}, \sdssa is absorbed by $N_{\rm H} = (8.1^{+2.9}_{-3.4}) \times 
10^{23}$~\nhunit.}
\label{bandratio_NH}
\end{figure}

\sdssa is detected with \nustar in the $8$--$24$~keV band, but not in the
$3$--$8$~keV band. We measure a
$99.7\%$ confidence level lower limit for the \nustar X-ray band ratio (i.e., the ratio of $8$--$24$~keV counts to $3$--$8$~keV
counts), of $>1.0$. In Figure
\ref{bandratio_z} we show the \nustar band ratio against redshift for \sdssa
and the first $10$ sources detected in the \nustar extragalactic
survey \citep{Alexander13}; the \sdssa band ratio is amongst the most
extreme.
We compare the band ratio with predictions from a simple absorbed power-law ({\sc
  zwabs$\cdot$pow}) model and the {\sc MYTorus} model \citep{Murphy09}, both of which are implemented in {\sc XSPEC}. 
{\sc MYTorus} is a self-consistent physical model that is valid for
the energy range $0.5$--$500$~keV, and for column densities of $N_{\rm
H}=10^{22}$--$10^{25}$~cm$^{-2}$. 
It is more suitable than the {\sc zwabs$\cdot$pow} model for column
densities of $N_{\rm
H}\gtrsim 5 \times 10^{23}$~\nhunit, where a careful
treatment of scattering and reflection is needed (for instance, see
Figure \ref{bandratio_NH}).
In the {\sc MYTorus} model, an obscuring torus reprocesses X-rays
from a central source, and the resulting X-ray spectrum has both transmitted and scattered components. 
In the current implementation of {\sc MYTorus}, the half-opening angle of the
obscuring medium is fixed to $60^{\circ}$ (i.e., a covering factor
of $0.5$), a value inferred from
the obscured AGN fraction of Seyfert galaxies. We note that a larger half-opening angle could be more appropriate in this
study of \typeii quasars, since the obscured AGN
fraction is observed to decrease with luminosity \citep[e.g.,][]{Ueda03,Lusso13}.
We assume a specific {\sc MYTorus} model with an intrinsic photon
index of $\Gamma = 1.8$ (typical
value for AGN at observed-frame $3$--$24$~keV; \citealt{Alexander13})
and an inclination angle of $\theta_{\rm obs}=70^{\circ}$, referred to
as {\it Model A} hereafter.
Varying $\theta_{\rm obs}$ between $65^{\circ}$ and $90^{\circ}$,
where $90^{\circ}$ corresponds to an edge-on
view through the equatorial plane of the torus,
makes a negligible difference to the {\sc MYTorus} band
ratio tracks in Figure \ref{bandratio_z}. We avoid using
$\theta_{\rm obs}$ values close to $60^{\circ}$, below which the line-of-sight
X-ray emission does not intercept the torus and the {\sc MYTorus} model
therefore describes an unobscured AGN. 
As shown in Figure \ref{bandratio_z}, the \nustar band ratio lower limit
for \sdssa corresponds to an
absorbing column density of $N_{\rm H} \gtrsim 2.5 \times
10^{23}$~cm$^{-2}$. This implies heavy, but not necessarily
Compton-thick, absorption. 

Since \xmm is more sensitive than \nustar at $< 8$~keV, we also
measure an X-ray band ratio for \sdssa using the \xmm $3$--$8$~keV data and \nustar
$8$--$24$~keV data, which gives a \nustar/\xmm band ratio of $1.2\pm
0.6$ ($68.3\%$ confidence level). 
One limitation of the measurement is that we are
unable to assess whether the X-ray emission of \sdssa has varied significantly 
in the $\sim 6.5$~years between the \xmm and \nustar
observations; if the \xmm count rate is relatively low, we 
overestimate the band ratio, and vice versa.
In Figure \ref{bandratio_NH}, we compare the measured \nustar/\xmm band ratio
with predictions from the {\sc MYTorus} and {\sc
  zwabs$\cdot$pow} models as a function of column
density. 
We fixed the model redshifts to that of \sdssa
($z=0.409$), used a range of intrinsic photon indices corresponding to those observed for unobscured AGN
\citep[$1.7<\Gamma<2.3$; e.g.,][]{Mateos10,Scott11}, and used a range of
inclination angles in the {\sc MYTorus} model ($65^{\circ}<\theta_{\rm obs}<90^{\circ}$). 
The resulting tracks in Figure
\ref{bandratio_NH} suggest that \sdssa is absorbed by $N_{\rm H} \gtrsim 5
\times 10^{23}$~\nhunit, which is consistent with the \nustar band ratio analysis (Figure \ref{bandratio_z}).
Assuming {\it Model A} ($\Gamma=1.8$ and $\theta_{\rm obs}=70^{\circ}$), the observed \nustar/\xmm band ratio for \sdssa implies a column density of $N_{\rm H} = (8.1^{+2.9}_{-3.4}) \times 
10^{23}$~\nhunit (i.e. heavy, but not clearly Compton-thick, absorption is
required to produce the observed $3$--$24$~keV X-ray spectrum).
This result is consistent with column density estimates from indirect
methods, as shown in Section \ref{results_indirect}.
For comparison, the highest column densities directly constrained by
\citet{Vignali06,Vignali10} in their $<10$~keV
analysis of SDSS-selected \typeii quasars are $N_{\rm H}\approx 3
\times 10^{23}$~\nhunit.

The above $N_{\rm H}$ constraint for \sdssa must be treated with a degree of
caution, since it depends on the assumed X-ray spectral model.
Here we assess the impact on our result of two spectral
complexities, both of which are important in the case of \typeii
quasars. First, a soft `scattered'
power law component is commonly observed for obscured AGN which may be either nuclear
emission scattered by hot gas \citep[e.g.,][]{Turner97}, or `leakage' of nuclear emission due to partial
covering \citep[e.g.,][]{Vignali98,Corral11}. %, or stellar
%emission \citep[e.g.,][]{Maiolino98b}. 
Adding a scattered component which is $2\%$ of the primary
transmitted power law \citep[a typical X-ray scattering fraction for
\typeii Seyferts; e.g.,][]{Turner97} to {\it Model A}, we obtain a consistent result: 
$N_{\rm H} > 4.9 \times 10^{23}$~\nhunit ($68.3\%$ confidence level lower limit).
Second, the absorbing medium may have a complex geometry (e.g.,
a clumpy torus) that requires the equatorial and line-of-sight column
densities of the {\sc MYTorus} model ($N_{\rm H,eq}$ and $N_{\rm H}$, respectively) to be treated
independently. Decoupling these two parameters in {\it Model A} and setting
$N_{\rm H,eq}$ to the maximum possible value of $10^{25}$~\nhunit
yields a consistent result: $N_{\rm H} = (7.7^{+2.8}_{-3.4}) \times
10^{23}$~\nhunit.
Last, 
we emphasize that although {\sc MYTorus} is a relatively complex
model, the $N_{\rm H}$ constraints do not differ significantly from those using a
simple {\sc zwabs$\cdot$pow} model in the Compton-thin regime (see Figure \ref{bandratio_NH}).
We conclude that the inferred $N_{\rm H}$ for \sdssa does not
change significantly with the assumed spectral model.

\subsection{Indirect Absorption Constraints}
\label{results_indirect}

\begin{figure*}
\centering 
\includegraphics[width=1\textwidth]{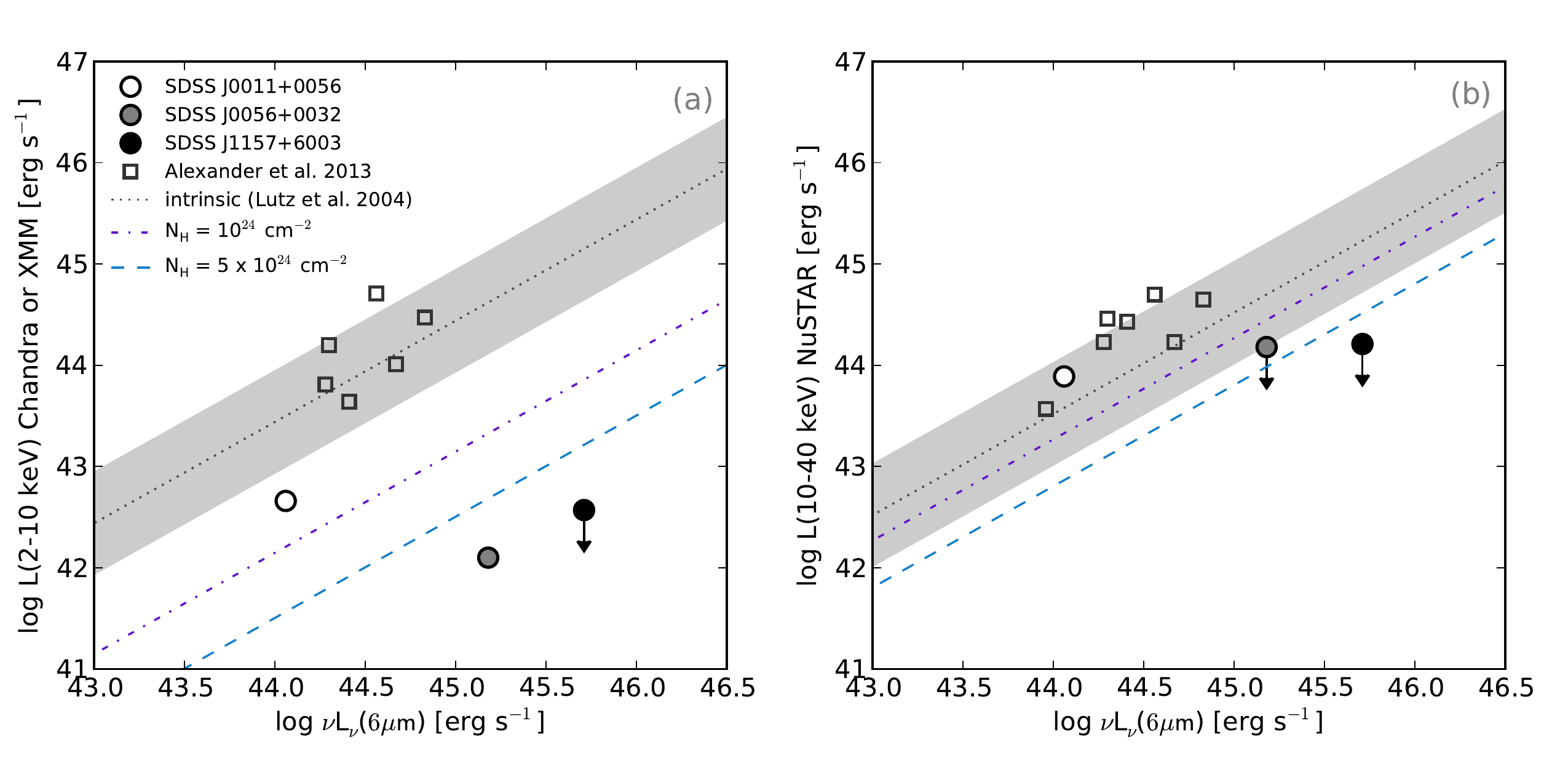}
\caption{Rest-frame X-ray luminosity against rest-frame
  \sixum luminosity for: (a) $2$--$10$~keV luminosities calculated
  using \xmm or \chandra data; and (b)
$10$--$40$~keV luminosities calculated using \nustar data. The X-ray
luminosities are not corrected for absorption. \sdssa,
\sdssb and \sdssc are shown as white, grey and black circles, respectively.
We compare with sources detected as part
of the \nustar extragalactic survey \citep[open squares;][]{Alexander13}. We also compare with an intrinsic relation
for $2$--$10$~keV, calibrated using local AGN \citep[dotted
line, with a shaded region indicating the scatter;][]{Lutz04}. 
This relation has been extrapolated to the
$10$--$40$~keV band assuming $\Gamma = 1.8$, and to relations for AGN absorbed
by $N_{\rm H} = 10^{24}$~cm$^{-2}$ (dash-dotted line) and $N_{\rm H} =
5 \times 10^{24}$~cm$^{-2}$ (dashed line) assuming a {\sc MYTorus} model with $\Gamma = 1.8$ and
$\theta_{\rm obs}=70^{\circ}$.
If we assume that low X-ray luminosities are due to absorption,
sources that lie below the $N_{\rm H} = 10^{24}$~cm$^{-2}$ tracks may
be Compton-thick.}
\label{LX_L6um}
\end{figure*}

The X-ray emission in heavily obscured AGN is subject
to significant absorption along the line of sight. The mid-IR emission, on the other
hand, has been reprocessed by the dust obscuring the AGN and is less
sensitive to extinction. The mid-IR luminosity therefore provides an estimate of the intrinsic AGN power. As such, the presence
of absorption in an AGN can be inferred
from the observed X-ray to mid-IR luminosity ratio
\citep[e.g.,][]{Lutz04,Alexander08,LaMassa09,Goulding11,LaMassa11}. We
note that the mid-IR emission is also significantly absorbed for $\approx 50\%$ of Compton-thick
AGN \citep[e.g.,][]{Bauer10,Goulding12}. Indeed, \sdssb has
significant Si-absorption at $9.7$~$\mu$m, in contrast to \sdssc (see
Section \ref{sample}). To account for this, we have corrected our
mid-IR luminosities for dust reddening (see Section \ref{UVMIR_data}).
In Figure \ref{LX_L6um} we compare the rest-frame X-ray luminosities ($L_{\rm X}$)
of our three \typeii quasars with the rest-frame \sixum luminosities
($L_{6\rm{\mu m}}$), exploring both the low energy ($2$--$10$~keV) and
high energy
($10$--$40$~keV) X-ray regimes. 
For \sdssa, $L_{\rm 2-10keV}$ was obtained through photometry in the
rest-frame $2$--$10$~keV band using \xmm data (see Section \ref{low_data}). For
\sdssb and \sdssc, $L_{\rm 2-10keV}$ was obtained through photometry in the observed-frame
$0.5$--$8$~keV band using \chandra data (see Section \ref{low_data}), and an extrapolation to the rest-frame
$2$--$10$~keV band assuming a power-law model with $\Gamma = 1.8$. The
$L_{\rm 10-40keV}$ values were obtained through a photometric
analysis in the rest-frame $10$--$40$~keV band using \nustar data (see Section
\ref{nustar_data}). The \sixum luminosities are from
SED fitting (Section \ref{UVMIR_data}) and relate specifically to the
emission from the AGN.

In the rest-frame $2$--$10$~keV band, the \typeii quasars fall below the intrinsic
X-ray--mid-IR luminosity relation found for AGN in
the local universe \citep{Lutz04}; see Figure
\ref{LX_L6um}{\it a}.
For comparison, we also show the non-beamed sources detected in 
the \nustar extragalactic survey \citep{Alexander13}, which lie within the
scatter of the \citet{Lutz04} relation.
The $2$--$10$~keV luminosity suppression of the three \typeii quasars is expected given our selection and has previously been
demonstrated for \sdssb and \sdssc \citep{Vignali06,Vignali10}.
Assuming the suppression of the X-ray emission is due to absorption, as opposed to
intrinsic X-ray weakness, we
estimate the column densities of these systems by comparing with the X-ray to mid-IR
luminosity ratios for AGN absorbed by $N_{\rm H} =
10^{24}$~cm$^{-2}$ and $N_{\rm H} = 5 \times 10^{24}$~cm$^{-2}$
(dash-dotted and dashed lines in Figure \ref{LX_L6um}{\it a}, respectively). 
On the basis of this analysis, the $2$--$10$~keV
luminosities of \sdssb and \sdssc are consistent with being absorbed by a factor
of $\gtrsim 300$, and therefore lie well within the Compton-thick
region with $N_{\rm H}\gtrsim 5 \times 10^{24}$~\nhunit. The X-ray
emission from \sdssa, on the other hand, is suppressed by a factor of
$\approx 7$, but is still consistent with being Compton-thick
or near Compton-thick ($N_{\rm H}\approx 10^{24}$~\nhunit).
Since our $2$--$10$~keV luminosities were calculated assuming
  a $\Gamma=1.8$ power-law, which is probably not consistent with heavy
  absorption at $z\sim0.5$, we repeated the flux
  calculations in Section \ref{low_data} assuming $\Gamma=0.6$ (the
  spectral slope of \sdssa as measured by \citet{Jia13}; see Section \ref{sample}).
This results in $L_{\rm 2-10keV}$ values which are higher by a factor
of $\approx 1.9$; not enough to significantly change the conclusions
drawn from Figure \ref{LX_L6um}{\it a}.

In the rest-frame $10$--$40$~keV band, the X-ray emission is only strongly suppressed for
column densities of $N_{\rm H} \gtrsim 5 \times 10^{24}$~cm$^{-2}$,
and therefore \nustar observes the intrinsic X-ray emission for all but
the most heavily obscured AGN; see Figure
\ref{LX_L6um}{\it b}.
%The relation for AGN absorbed by
%$N_{\rm H} = 10^{24}$~cm$^{-2}$ is almost equivalent to that for unobscured AGN at these
%high X-ray energies.
For comparison, \citet{Matsuta12} studied {\it Swift}/BAT-detected AGN and
found that for $14$--$195$~keV, only $\approx 60\%$ of Compton-thick objects have
significant X-ray suppression with respect to the intrinsic X-ray to
mid-IR luminosity ratio.
The results in Figure \ref{LX_L6um}{\it b} suggest that the X-ray
emission from \sdssa is not significantly suppressed at $10$--$40$~keV, and is 
absorbed by $N_{\rm H} \lesssim 10^{24}$~\nhunit. This is consistent
with the X-ray band ratio analysis in Section \ref{results_direct}. 
\sdssb is consistent with being Compton-thick,
with $N_{\rm H} \gtrsim 10^{24}$~\nhunit. 
\sdssc is the strongest candidate for being Compton-thick based on
this analysis. Its $10$--$40$~keV luminosity is consistent with being
absorbed by a factor of $\gtrsim 10$, despite the high X-ray energies
being probed, which again suggests an extreme column density of $N_{\rm H}
\gtrsim 5 \times 10^{24}$~\nhunit. 
Assuming $\Gamma=0.6$, rather than $\Gamma=1.8$, for the
  \nustar count rate to flux conversion (Section \ref{flux_calc}) results in $L_{\rm 10-40keV}$ values which
  are a higher by a factor of $\approx 1.4$; again, not enough to
  significantly change the conclusions drawn from Figure \ref{LX_L6um}{\it b}.
As an independent test, we repeated our indirect analyses using \oiii luminosity as a measure of
intrinsic AGN power (i.e., using $L_{\rm X}/L_{\rm [OIII]}$). This yielded very similar results; 
\nustar observes the intrinsic X-ray emission of
\sdssa, while \sdssb and \sdssc are consistent with being heavily Compton-thick
($N_{\rm H} \gtrsim 5 \times 10^{24}$~\nhunit). 
However, since our sample was originally
selected on the basis of high \oiii luminosity \citep{Zakamska03,Reyes08}, we consider the $L_{\rm X}/L_{\rm 6 \mu m}$ results to be more reliable.
Nevertheless, the $L_{\rm X}/L_{\rm 6 \mu m}$ ratio alone is not a robust
indicator of Compton-thick absorption, even if the \sixum emission accurately reflects the intrinsic power of
the AGN. 
First, some quasars can be intrinsically X-ray weak (e.g.,
\citealt{Gallagher01,Wu11,Luo13}; Teng
et al. 2013, ApJ, submitted).
Second, inferred column densities
depend on the assumed X-ray spectral model \citep[e.g.,][]{Yaqoob11,Georgantopoulos11b}.
For instance, adding an additional soft scattered 
component, with a scattering fraction of $2\%$, to the {\sc MYTorus} model predicts a $L_{\rm 2-10keV}/L_{\rm 6 \mu m}$ ratio for $N_{\rm H} = 5
\times 10^{24}$~\nhunit which is a factor of three
higher than that shown in Figure \ref{LX_L6um}{\it b}. However, this is not enough to
change our broad conclusions regarding the column densities of \sdssb and \sdssc.
Ultimately, deeper X-ray observations, with simultaneous coverage at
low and high energies, are required to
directly constrain $N_{\rm H}$ and provide more
robust evidence for or against the presence of Compton-thick absorption
in these \typeii quasars.

%\input{discussion/discussion.tex}

%%%%%%%%%%%%%%%%%%%%%%%%%%%%%%%%%%%%%%%%%%%%%%%%%%%%%%%%%%%%%%%%%%%%%%
\section{Summary and Future Work}
\label{summary}
%%%%%%%%%%%%%%%%%%%%%%%%%%%%%%%%%%%%%%%%%%%%%%%%%%%%%%%%%%%%%%%%%%%%%%

We have presented the first sensitive high energy ($>10$~keV) analysis
of optically selected \typeii quasars. The sample consists of three objects
that show evidence for Compton-thick absorption ($N_{\rm H}>1.5\times 10^{24}$~\nhunit) on the basis
of different diagnostics (see Section \ref{sample}). To summarize our main results:

\begin{itemize}

\item One of the \typeii quasars, \sdssa, is detected by \nustar with
  $16.8 \pm 6.4$ counts in the $8$--$24$~keV band. The remaining two,
  \sdssb and \sdssc, are not detected by \nustar; see Section \ref{photometry}.

\item For \sdssa, we characterize the $3$--$24$~keV
  spectrum using the X-ray band ratio and find evidence for near
  Compton-thick absorption with $N_{\rm H} \gtrsim 5 \times 
10^{23}$~\nhunit; see Section \ref{results_direct}. 
This is consistent with the column densities inferred from the
$2$--$10$~keV to mid-IR ratio, the $10$--$40$~keV to mid-IR ratio, and the
X-ray to \oiii ratios; see Section \ref{results_indirect}.

\item For \sdssb and \sdssc, we find evidence for a significant
  suppression of the rest-frame $10$--$40$~keV luminosity with respect to
  the mid-IR luminosity. If due to absorption, this result
  implies that these \typeii quasars are extreme, Compton-thick
  systems with $N_{\rm H} \gtrsim 10^{24}$~\nhunit; see Section \ref{results_indirect}.

\end{itemize}

The characterisation of distant
heavily obscured AGN is clearly an extremely challenging pursuit.
Nevertheless, as we have demonstrated, the sensitive high energy
observations of \nustar provide a significant improvement compared to
\chandra or \xmm observations alone; for quasars at $z\sim0.5$, high column densities
  of $N_{\rm H}\gtrsim 5 \times 10^{23}$~\nhunit can now be directly
  constrained.
Based on the results obtained in this exploratory study, we are now extending the analysis of optically
selected \typeii quasars to a larger sample which is currently being observed by \nustar.
Furthermore, \nustar is undertaking deep surveys in the ECDFS (Mullaney et al., in prep.)
and COSMOS (Civano et al., in prep.) fields, along with a large-area serendipitous
survey \citep{Alexander13}, that are likely to uncover a number of heavily obscured quasars.
These upcoming studies will provide a leap forward in our understanding of the column
density distribution of distant luminous AGN.

%%%%%%%%%%%%%%%%%%%%% acknowledgements %%%%%%%%%%%%%%%%%%%%%%%%%%%%%%%
\section*{Acknowledgements}

We acknowledge financial support from the Science and Technology
Facilities Council (STFC) grants ST/K501979/1 (G.B.L.), ST/I001573/1 (D.M.A. and A.D.M.) and ST/J003697/1 (P.G.), the Leverhulme
Trust (D.M.A. and J.R.M.), Gemini-CONICYT grant 32120009 (R.J.A.), NSF AST award 1008067 (D.R.B.), the International Fulbright Science and Technology Award (M.B.), Basal-CATA PFB-06/2007 (F.E.B.),
CONICYT-Chile grant FONDECYT 1101024 (F.E.B.), CONICYT-Chile grant
Anillo ACT1101 (F.E.B.), Caltech \nustar subcontract
44A-1092750 (W.N.B. and B.L.), NASA ADP grant NNX10AC99G (W.N.B. and
B.L.), NASA ADAP award NNX12AE38G (R.C.H.), National Science Foundation grant 1211096 (R.C.H.), and Swiss National Science Foundation grant PP00P2\_138979/1
(M.K.). We thank the referee for the constructive comments, which helped improve our study.
This work was supported under NASA Contract No.\ NNG08FD60C, and made use of data from the \nustar mission, a project led by the California Institute of Technology, managed by the Jet Propulsion Laboratory, and funded by the National Aeronautics and Space Administration. We thank the \nustar Operations, Software and Calibration teams for support with the execution and analysis of these observations. This research has made use of the \nustar Data Analysis Software (NuSTARDAS) jointly developed by the ASI Science Data Center (ASDC, Italy) and the California Institute of Technology (USA).
%%%%%%%%%%%%%%%%%%%%%%%%%%%%%%%%%%%%%%%%%%%%%%%%%%%%%%%%%%%%%%%%%%%%%%

%\input{appendix/appendix.tex}

%%%%%%%%%%%%%%%%%%%%%%%%%%%%%%%%%%%%%%%%%%%%%%%%%%%%%%%%%%%%%%%%%%%%%%

%%%%%%%%%%%%%%%%%%%%%%%%%%%%%%%%%%%%%%%%%%%%%%%%%%%%%%%%%%%%%%%%%%%%%%
\end{document}